\documentclass[reprint,aps,amsmath,amssymb,nofootinbib]{revtex4-1}
\usepackage{amssymb}
\usepackage{amssymb}
\usepackage{amssymb}
\usepackage{graphicx}
\usepackage{graphics}
\usepackage{amsmath}
\usepackage{placeins}
\usepackage{hyperref}
\usepackage[dvipsnames]{xcolor}
\usepackage{wasysym}
\usepackage{soul}
\usepackage{float}
\usepackage{array}
\usepackage{tabulary}
\usepackage{caption}
\usepackage{fancyhdr}
\usepackage{epstopdf}

\begin{document}
	
	\title{Evidence for Anti-Dowell-Schmerge Process in Photoemission from Diamond}
	
	\author{Tanvi Nikhar}
	\author{Sergey V. Baryshev}
	\email{serbar@msu.edu}
	\affiliation{Department of Electrical and Computer Engineering, Michigan State University, 428 S. Shaw Ln., East Lansing, Michigan 48824, USA}
	
    \author{Gowri Adhikari}
    \author{Andreas W. Schroeder}
    \email{andreas@uic.edu}
    \affiliation{Department of Physics, University of Illinois at Chicago, 845 W. Taylor St., Chicago, Illinois 60607, USA}
    
	
\begin{abstract}
A great number of metal and semiconductor photocathodes, which are of high practical importance for photoinjector applications, closely follow the 1/3 gradient Dowell-Schmerge (DS) law describing the spectral dependence of the mean transverse energy ($MTE$), $viz.$ $MTE$ as a function of the incident laser photon energy. However, some (rare) semiconductor photocathodes show $MTE$ trends that are significantly different. For example, spectral $MTE$ measurements on PbTe, BaLaSnO or Hf/HfO$_2$ have clearly demonstrated trends that can differ from DS law being non-monotonic, slower growing, or displaying constant $MTE$ versus laser photon energy. We have discovered that $n$-type ultra-nano-crystalline diamond (UNCD) and single crystal diamond are anti-DS photocathodes in that their $MTE$ decreases with the incident photon energy. It was previously established that UNCD is a highly emissive material in the near UV such that quantum efficiency ($QE$) grows with the laser photon energy. The unique and novel combination of high increasing $QE$ and low decreasing $MTE$ of UNCD may pave the way to desired high brightness electron beams, through operation well above its work function which fundamentally differs from ‘Boltzmann tail’ operation near the photoemission threshold. One other remarkable result followed: As UNCD is a $sp^2$ grain boundary diluted $sp^3$ diamond matrix, control over grain boundary/grain engineering in the material’s synthesis allowed for the production of different kinds of UNCD. The resultant tuning of the $sp^3$-to-$sp^2$ ratio in different UNCD photocathodes allowed for switching between canonical +1/3 DS and approximate --1/3 gradient ‘anti-DS’ behavior.
\end{abstract}

\maketitle

Photocathode-based rf and pulsed dc guns are bright electron injectors for light sources and advanced time-resolved microscopes. The further progress of X-ray and microscopy facilities (improved sensitivity, spatio-temporal resolution, and high throughput) will largely depend on the understanding and consequent development of materials with the potential to be utilized as photocathodes. The ratio of the photo-emitted charge to the mean transverse energy ($MTE$) of the emitted electrons determines the photocathode brightness, which in many applications is the most critical figure of merit. In these terms, the photocathode brightness $B$ can be defined as $B=\frac{2m_e c^2 I}{\sigma_x^2 MTE}$ or $\propto \frac{QE}{MTE}$, where $I$ is the peak beam current, $QE$ is quantum efficiency, $\sigma_x$ is the adjustable r.m.s. laser spot size on the photocathode, and the electron rest mass $m_e$ and speed of light $c$ are fundamental constants. As the emitted beam current density $I⁄(\sigma_x^2 )$ can be increased up to the optical damage (or melting) threshold of the photocathode surface, it is the $MTE$ -- an intrinsic material property -- that must be reduced to achieve the high brightness goals and that have become central to photocathode R\&D.

For a classical metal photocathode such as copper at zero temperature (0 K), Dowell and Schmerge (DS) showed that the $MTE$ increases linearly as $MTE=\frac{1}{3}\Delta E$ with the excess photoemission energy $\Delta E=\hbar\omega-\phi$, where $\hbar\omega$ is the primary incident laser photon energy and $\phi$ is the photocathode work function. At higher temperatures, further work has shown that for $\Delta E < 0$ (below work function, sub-threshold, photoemission) the limiting minimum value for the $MTE$ (due to emission from the Boltzmann tail of the electron distribution in the photocathode material) is about $k_B T_e$, where $T_e$ is the electron temperature and $k_B$ is Boltzmann’s constant. For these reasons, to attain the highest quality (low divergence) electron beam, metal photocathodes are often operated in the near threshold region; that is, at a small $\Delta E$ with the primary photon energy nearly matching the work function. Operating in this ‘Boltzmann tail’ regime requires larger laser power density to enable high charge injection. This may lead to emittance growth due to the laser-heating of the photocathode that may become a problem at high repetition rates -- an effect that can be minimized through the proper choice of high $QE$ material or by operating the photocathode at primary laser energy well above the work function (also increasing the $MTE$) as both strategies would increase primary photon-to-secondary photoelectron conversion efficiency.

In our previous work, the $MTE$ of UNCD photocathode was evaluated using primary wavelengths from 4.4 to 5.3 eV \cite{MTEAPL2019}. Interestingly, the $MTE$ demonstrated no noticeable dependence on the photon energy, with an average value of 266 meV. This spectral behavior was shown not to be dependent upon physical or chemical surface roughness, and was inconsistent with low electron effective mass emission from graphitic grain boundaries. The major hypothesis set forth was that it was due to UNCD’s unique nanoscopic periodic $sp^3$ grain / $sp^2$ grain boundary (GB) structure. It was hypothesized that at low photon energies the emitted electrons originate from spatially confined ground states in the graphitic GB’s (because they carry electrons) between the diamond grains. If so, transverse momentum conservation in photoemission implies that the $MTE$ should reflect the r.m.s. momentum (i.e., size in momentum space) of these states. Through the Heisenberg’s uncertainty principle $\Delta p\Delta x\sim h$, one can roughly estimate that the obtained $MTE$ of $\sim$200 meV corresponds to a spatial region of $\sim$1 nm, which was in good agreement with the GB size in UNCD. Following this original hypothesis, we synthesized a new set of UNCD photocathode thin films in which $sp^3$ to $sp^2$ ratio was greatly varied in the effort to observe the $MTE$ scaling with GB size.

Unlike single crystal diamond, UNCD contains graphitic, or $sp^2$ hybrized, GBs, making UNCD a two-phase material comprised of $sp^3$ diamond grains ($\sim$10 nm in size) and $sp^2$-rich graphitic GBs ($\sim$1 nm in size). The graphitic GBs provide the charge to compensate the surface while the diamond grains can be functionalized with hydrogen or other dipole allowing for work function engineering -- all critical for photocathode design. UNCD growth can be achieved through various chemical vapor deposition (CVD) processes. Regardless of the process [whether it is H$_2$-rich or Ar-rich microwave (H$_2$ or Ar)/CH$_4$ plasma assisted or hot filament CVD], it has been found that $sp^3$-to-$sp^2$ ratio can be tuned through 1) incorporating nitrogen into the GBs by adding N$_2$ or ammonia gas to the precursor gas mixture, 2) adjusting the deposition temperature $T_D$, or 3) changing CH$_4$ in H$_2$-rich plasma -- all give nearly same effect on conductivity \cite{Zapol,Birrell2002,Orlando2016,0N2,uwave}. The UNCD films in this study were grown on ultrasonically seeded intrinsic Si(100) substrates using a microwave plasma CVD technique in a reactor operated at 2.45 GHz \cite{ASMUSSEN2012}. Samples were grown from a hydrogen rich H$_2$/N$_2$/CH$_4$ precursor gas mixture with different nitrogen volume \% (varied between 0 and 20\%), which corresponded to flow rates of 0-40 standard cubic centimeters per min (sccm) of N$_2$. The CH$_4$ flow rate was kept constant at 10 sccm (5\% by volume), while the flow rate of N$_2$ and H$_2$ was varied to maintain a total flow rate of 200 sccm. Different deposition temperature (equal to measured substrate temperature) $T_D$ were achieved, ranging 1043 to 1295 K, by varying the total pressure of the vacuum chamber (35 to 60 torr) and the input microwave power (2.5 to 3.0 kW). The substrate temperature was measured using an infrared pyrometer during growth process of 60 minutes for each sample. Ref.\cite{Tanviarxiv} provides in-depth analyses of UNCD structure upon its synthesis.

\begin{figure}[h]
\begin{center}
\includegraphics[height=7.cm]{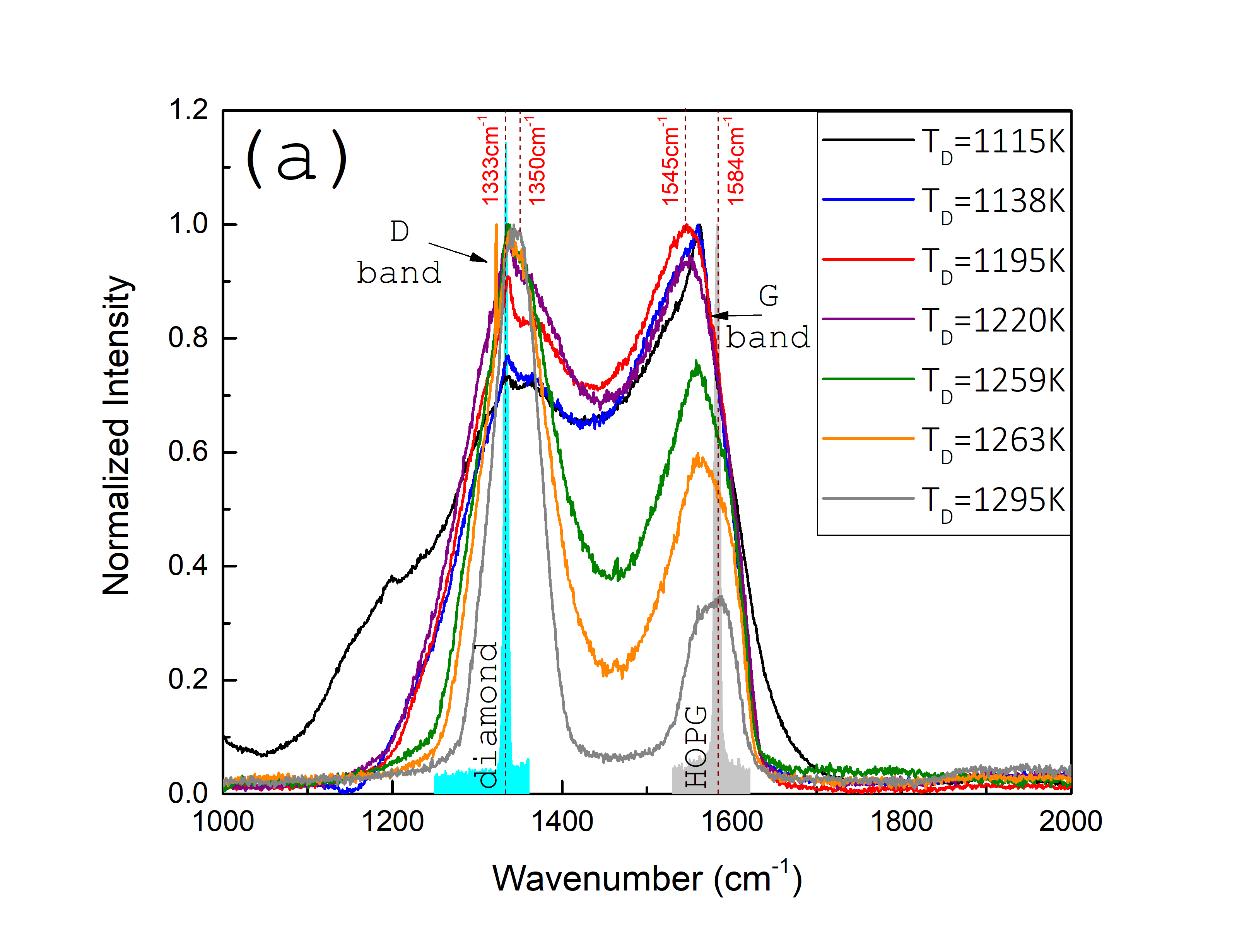}
\includegraphics[height=7.cm]{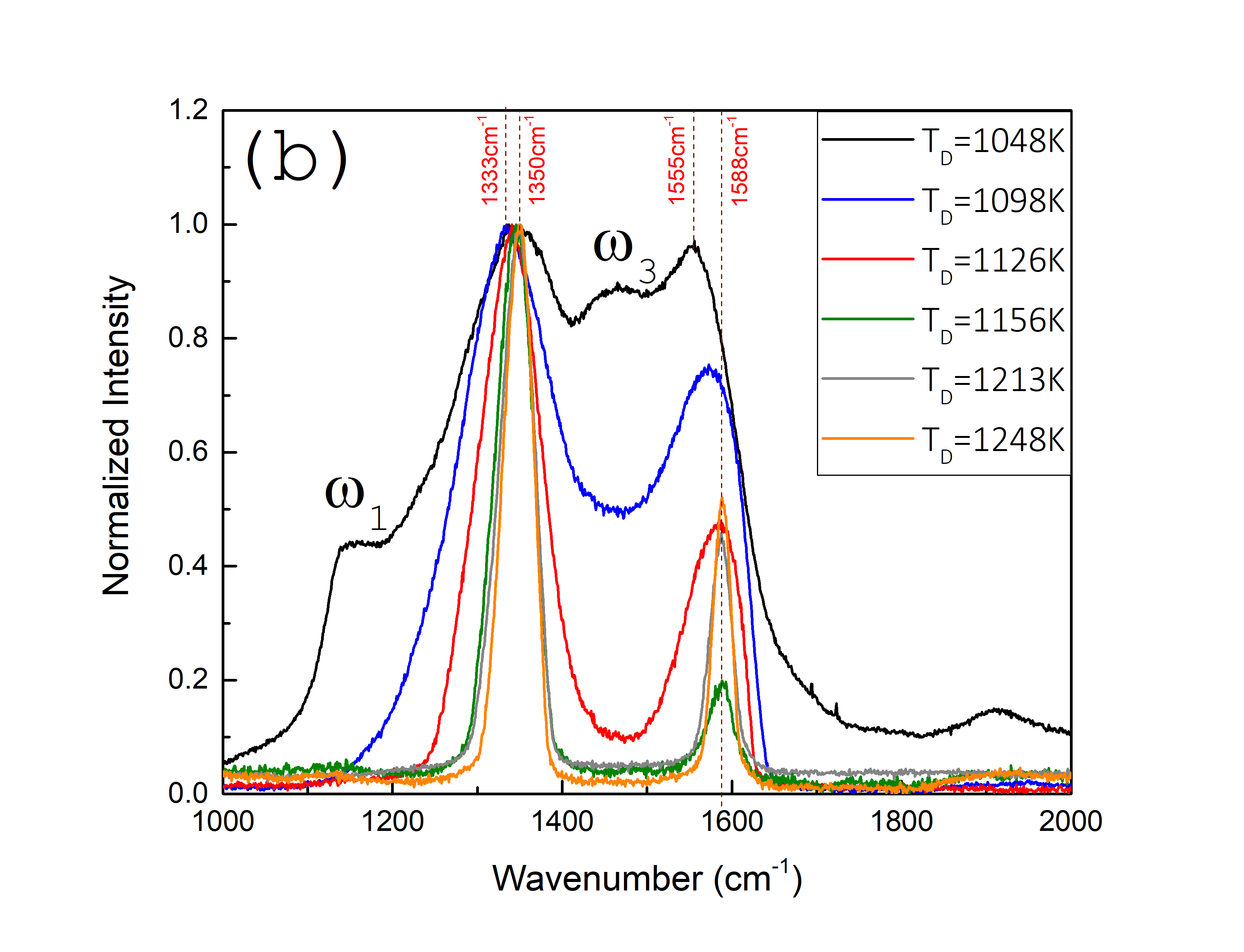}
\caption{Raman spectra for (a) 20\% and (b) 0\% N$_2$ samples as a function of $T_D$.}
\label{raman}
\end{center}
\end{figure}

\begin{figure*}[!]
\begin{center}
\raisebox{0.1\height}{\includegraphics[height=4.5cm]{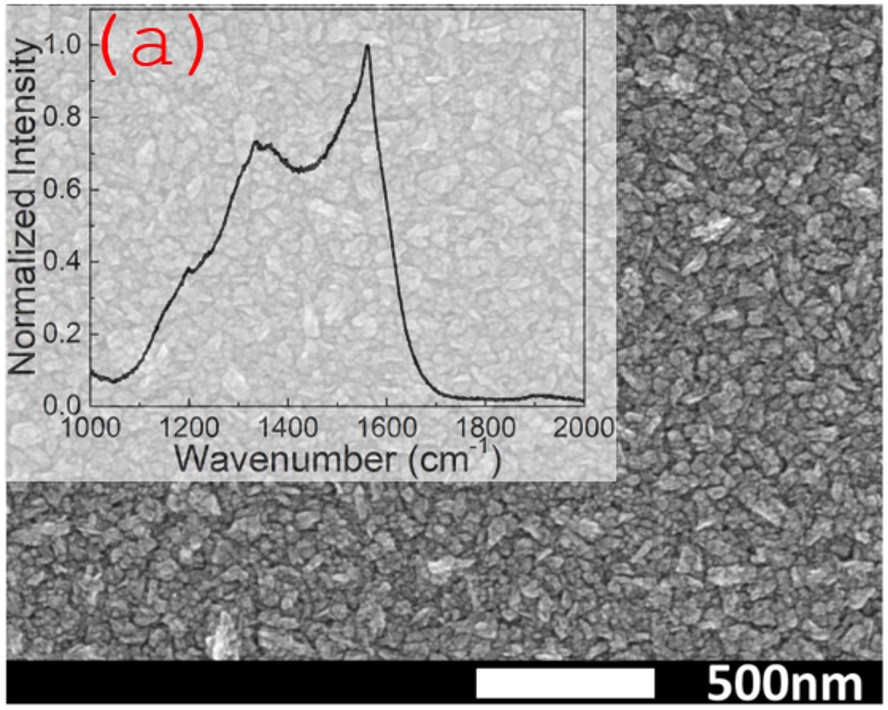}} \raisebox{0.1\height}{\includegraphics[height=4.5cm]{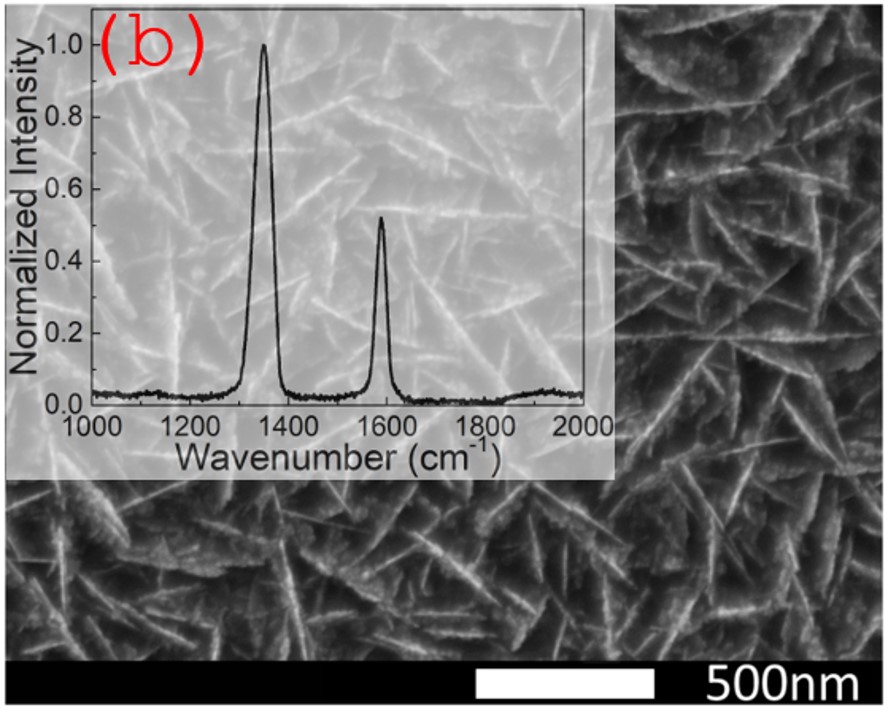}} \includegraphics[height=5.cm]{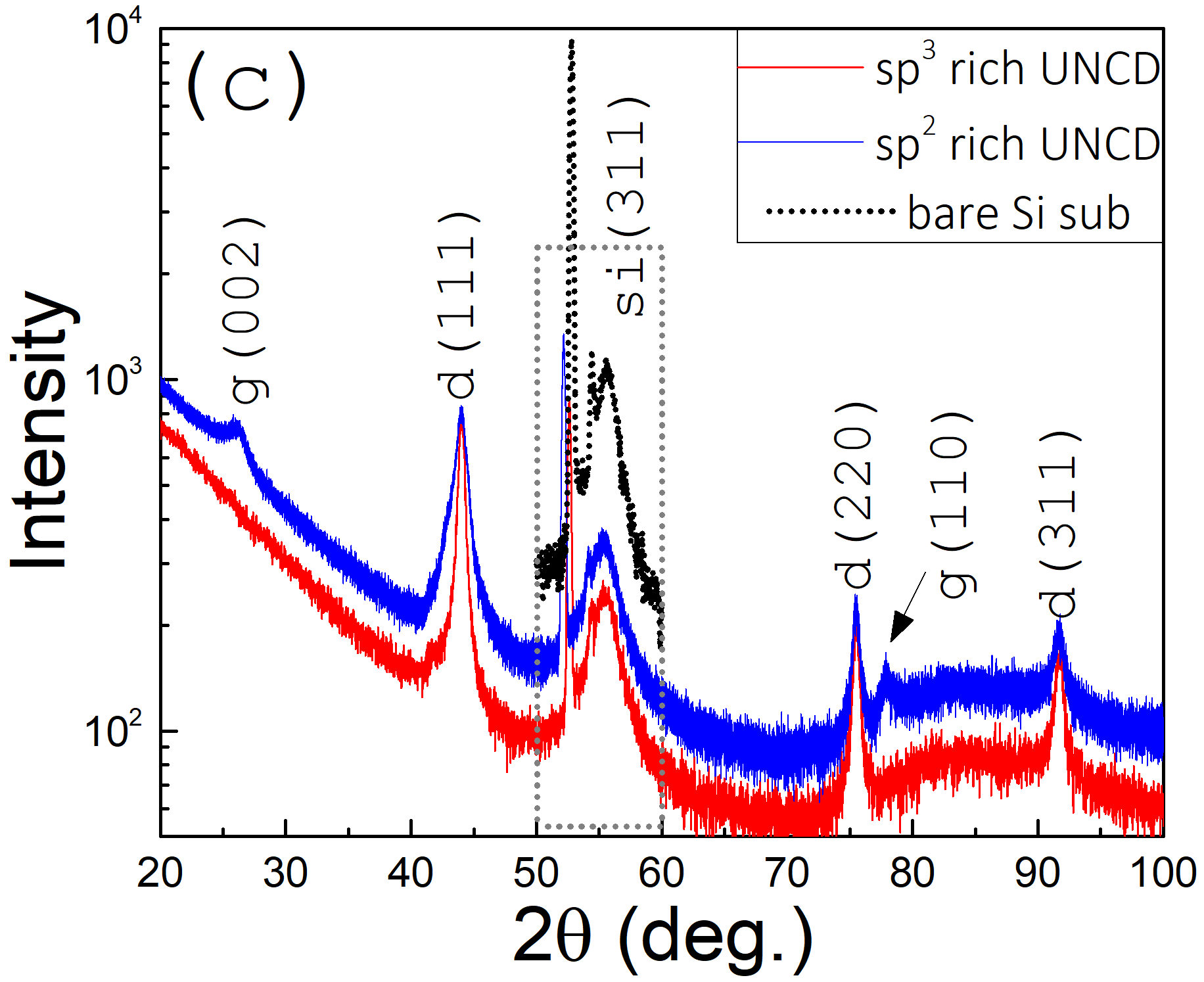}
\caption{Top view SEM of (a) $sp^3$-rich UNCD grown with 20\% N$_2$ at 1115 K and (b) $sp^2$-rich UNCD grown with 0\% N$_2$ at 1248 K. (c) Grazing incidence XRD spectra showing graphite peak formation as UNCD transforms from nanodiamond and nanographite state; it also shows domination of diamond 111 phase: $d$ stands for diamond and $g$ stands for graphite peaks.}
\label{sem}
\end{center}
\end{figure*}

A Raman spectroscopy summary for a set of samples grown at different $T_D$ is presented in Fig.~\ref{raman}. Two bands labeled D and G are observed. The D band is located between 1333 and 1350 cm$^{-1}$. The 1333 cm$^{-1}$ peak is the characteristic diamond peak attributed to $sp^3$ bonded carbon. The 1350 cm$^{-1}$ feature appears for graphitic materials in the form of a band induced by the defect or disorder in the $sp^2$ bonded carbon. For instance, polycrystalline graphite exhibits this D band contribution in its Raman spectrum due to the outer defective rim of graphite crystallites. These two spectral features of the D band are well-resolved for low $T_D$ 20\% nitrogen films (the red 1195 K curve in Fig.~\ref{raman}a): one sharp feature centers at 1333 cm$^{-1}$ and is interpreted to be due to $sp^3$ phase, while the second one centers at near 1350 cm$^{-1}$ and is due to disordered $sp^2$ bonded carbon. Since resolved diamond peak at 1333 cm$^{-1}$ is rarely seen in UNCD films, it is clear that our synthesis extends into nano-diamond class \cite{mrsbull}. This is confirmed by SEM: Fig.~\ref{sem}a illustrates surface consisting of large faceted diamond grains. The second major spectral feature in Fig.~\ref{raman} is a peak ranged between 1545 and 1590 cm$^{-1}$, called the G band. While top curves in Fig.~\ref{raman}a and b with their G peak positioned at 1550 cm$^{-1}$ suggest that $sp^2$ grain boundaries are largely in an amorphous state, the bottom curves (for progressively larger $T_D$) with its G peak positioned at 1590 cm$^{-1}$ suggests that well-crystallized nanographite forms in the grain boundaries \cite{ferrari}. Simultaneously, the D peak moves from 1333 to 1350 cm$^{-1}$ and changes its meaning -- D for diamond to D for defect (nanographite defects) \cite{dress}. At temperatures $\sim$1200 K (depending on amount of N$_2$ in plasma), UNCD heavily graphitizes, and the diamond phase vanishes per the Raman analyses. X-ray diffraction also shows graphite formation in $sp^2$ rich UNCD [blue line in Fig.~\ref{sem}c with emerging graphite peaks (002) and (110)]. Nanographite platelets are then formed as seen by SEM (Fig.~\ref{sem}b), and material transformes from nano-diamond to nano-graphite. A further basic confirmation of this trend comes from comparing SEM film thickness measurements against sample weighing: it was found that the mass density of UNCD changes from 3.51 g/cm$^3$ (diamond-like) to 2.27 g/cm$^3$ (graphite-like).

Three types of photocathode samples were studied: 1) nanodiamond (low $T_D$ 20\% N$_2$ UNCD demonstrating resolved sharp diamond peak at 1333 cm$^{-1}$, $sp^2$ content $\sim$1 at.\%); 2) nanographite (high $T_D$ 0\% N$_2$ UNCD, producing heaviest graphitization, with D peak shifted to 1350 cm$^{-1}$ and G peak shifted to 1590 cm$^{-1}$, $sp^2$ content $>>$10 at.\%); and 3) canonical UNCD (mid-$T_D$ 10\% N$_2$ UNCD, $sp^2$ content $\sim$10 at.\%). By ‘canonical UNCD’ we mean a sample for which the D and G Raman bands are prominent: the D peak is wide because the diamond and graphite defect D peaks are not resolved from each other, and the G peak remains strong, while the $\omega_1$ and $\omega_3$ bands (best visible in Fig.~\ref{raman}b) reflect the presence of hydrocarbon transpolyacetylene chains contained in GBs. Two reference samples were also used: 1) pure $sp^2$ hybridized highly oriented pyrolytic graphite (HOPG) to mimic nanographite $sp^2$-rich form of UNCD; and 2) pure $sp^3$ single-crystal diamond (111) grown by the high pressure high temperature (HPHT) method with high nitrogen concentration.  For the latter, the nitrogen content and crystal orientation were chosen to mimic nanodiamond form of UNCD. The Raman spectra of HOPG and diamond are shown in Fig.~\ref{raman}a.

The spectral $MTE$ measurements of the photocathode samples employ a tunable ultraviolet (UV) radiation source driven by a 30 MHz repetition rate, diode pumped, mode-locked Yb:KGW oscillator \cite{was1} to generate electron pulses in a 20 kV dc gun \cite{was2}. The sub-picosecond UV radiation source is produced using optical parametric amplification of a nonlinear fiber generated continuum followed by sum-frequency generation to obtain near continuous tunability from 3.0-5.3 eV (235-410 nm). The procedure employed to measure the $MTE$ of the electrons photoemitted from the photocathodes has been reported elsewhere \cite{was3}. Briefly, the solenoid scan technique \cite{was4} is used to measure the electron beam spot size on a Ce:GAGG \cite{GAGG} scintillator screen (1:1 image onto a CCD camera) as a function of the focal strength of the two counter-wound cylindrical (solenoid) magnetic lenses. The $MTE$ is then determined by fitting the measured trend with an extended Analytical Gaussian (AG) electron pulse propagation model \cite{was4, was5,was6}, whose input parameters include the r.m.s. laser spot size on the photocathode and the 3D acceleration field distribution in the 20 kV dc gun.

\begin{figure}[h]
\begin{center}
\hspace*{-0.7cm}\includegraphics[height=8cm]{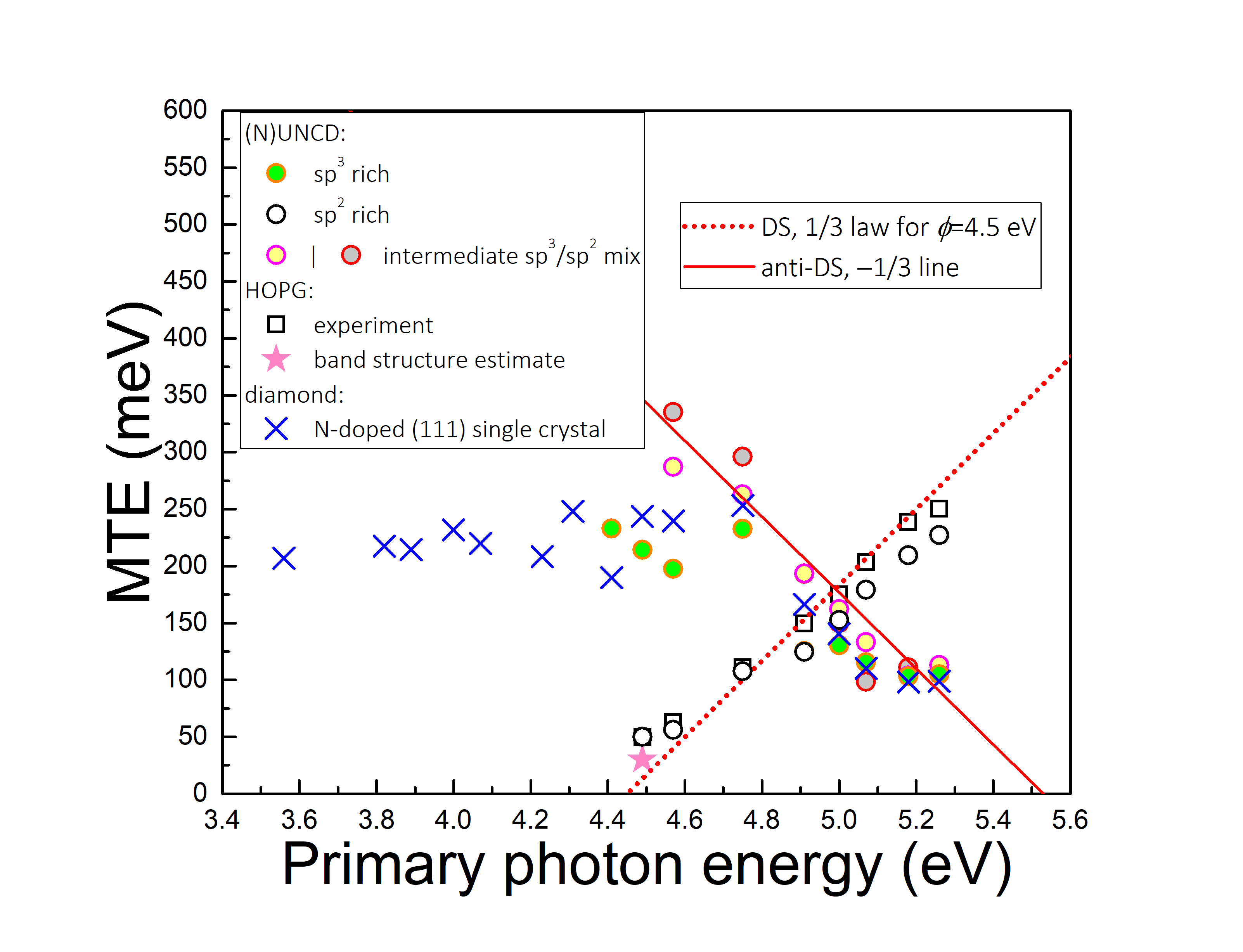}
\caption{Experimental demonstration of anti-DS UNCD photocathode: $MTE$ values are plotted versus the primary incident photon energy. The $MTE$ data is compared against pure $sp^2$-hybrized highly oriented pyrolytic graphite (HOPG) that appears to be a classical DS photocathode (DS law is depicted by the red dotted line) and against pure $sp^3$-hybrized diamond that featured unconventional behavior. Reversed, negative slope, anti-DS $MTE$ trend is also shown by the red solid line.}
\label{mte}
\end{center}
\end{figure}

Fig.~\ref{mte} presents the full summary of $MTE$ values obtained for the six photocathode samples using multiple solenoid scans performed at multiple primary laser photon energies. Remarkably, the measured $MTE$ values display a negative slope trend for the intermediate $sp^3$/$sp^2$ mix of canonical UNCD (sample type 3), rolling off from 335 to 98 meV with the UV photon energy rising from 4.57 to 5.26 eV (red open and gray-filled circles). Moreover, no anticipated $MTE$ scaling with GB size was observed. Instead, a new photoemission mechanism was discovered. Results obtained for $sp^3$-rich nanodiamond (sample type 1) further confirmed this unusual trend, with $MTE$ reaching down at 98 meV (green-filled circles). One difference exists -- the observed emission threshold is lower, 4.4 eV, and the $MTE$ is constant at about 215 meV before it rolls off. Comparison with the nitrogen doped single crystal diamond (111) (blue crosses) implies that the flat $MTE$ response must be due to defect-mediated emission from nitrogen centers contained in $sp^3$ grains of nanodiamond. In fact, an even earlier emission threshold at 3.56 eV was observed for nitrogen doped diamond, and $MTE$ remained constant at $\sim$215 meV until the laser energy reached 4.75 eV. Thereafter, the $MTE$ follows exactly the values for $sp^3$-rich and intermediate $sp^3$/$sp^2$ mix samples -- all fit onto a \mbox{--1/3} gradient anti-DS line (solid red line) that extends to higher $MTE$ values at lower photon energies that also captures the type 3 sample behaviour.

An even more remarkable finding is that UNCD can also show the classical DS behavior upon $sp^2$ phase engineering. In graphitized or $sp^2$-rich UNCD, a short range order semi-amorphous conducting system of GB (as in canonical UNCD) becomes well crystallized (a long-range order). Consequently a prediction can be made that nanodiamond converted into nanographite should lose the spatially confined photoemission uniqueness, and so the $MTE$ would switch back to canonical DS, $\frac{\hbar\omega-\phi}{3}$, law (highlighted with the red dotted line in Fig.~\ref{mte} for work function $\phi$=4.5 eV.) From Fig.~\ref{mte}, one can see that spectral dependence of the $MTE$ did switch from --1/3 anti-DS to +1/3 DS line for the $sp^2$-rich photocathode sample (open black circles). In fact, nanographite-like UNCD behaves exactly like the reference single crystal HOPG (a ‘perfect’ DS photocathode, see open black squares).

In unusual way, the presented results may be looked at as confirming our original hypothesis: transverse momentum conservation in photoemission implies that the $MTE$ should reflect the r.m.s. momentum (i.e., size in momentum space) of the emission states. Through the Heisenberg’s uncertainty principle ($\Delta p\Delta x\sim h$), one can roughly estimate that the obtained $MTE$ of $\sim$180 meV at the crossing point between DS and anti-DS trend lines in Fig.~\ref{mte} corresponds to a spatial emission size of $\sim$10 nm, which is in good agreement with the grain size in our UNCD photocathode samples. The further drop in the $MTE$ with increasing photon energy may imply that photoemission creeps out toward larger diamond grains as the phase space volume of $\sim h$ is conserved. A further supporting argument is that the DS and anti-DS lines cross at about 5 eV -- a photon energy near the indirect band gap of diamond. We note that since no surface charging effect was observed during long experimental runs, the GB network in the UNCD photocathode replenishes the photoemitted charge.

The explanation of the threshold nature of the photoelectric effect given by Einstein, the Fowler and DuBridge law relating the $QE$ to the primary photon energy and material’s work function, and the DS law relating the transverse electron momentum (or $MTE$) to the primary photon energy established a basis for the cathode R\&D with its practical design paradigms and limitations. Diamond and nanodiamond photocathodes demonstrate the expected $QE$ growth with incident photon energy (like in Fowler-DuBridge law) \cite{diamondQE,UNCDQE}, but violate the Dowell-Schmerge law by demonstrating a linear $MTE$ decrease with incident photon energy. This means that new bright cathode designs become feasible, allowing low intrinsic emittance (or $MTE$) operation well above the emission threshold (well above the work function) rather than in the Boltzmann tail regime. Operation at primary photon energies at or above 5 eV, where the $QE$ increases and the $MTE$ decreases, should minimize thermally induced emittance growth and maximize the brightness of the photocathode placed in the properly designed gradient and triggered at the properly designed phase of an rf photo-electron gun.

To conclude, the observed anti-DS behavior is possibly a unique property of diamond. Fortunately, nanodiamond and UNCD retain this property which makes this material a unique photocathode platform because it is also conductive and hence immune to surface charging effects via facile and ultrafast electron transport through graphite-like GB networks. Further DFT analysis would be required to provide insights as to what role various conduction band valleys within the diamond band structure play to produce such a novel $MTE$ behaviour. Together with the HOPG data, DFT could provide a baseline for the understanding of the electronic states in various UNCD samples and thus the DS / anti-DS switching induced by the interplay between $sp^3$ and $sp^2$.

\

\noindent \textbf{Acknowledgment.} T.N. and S.V.B. were supported by funding
from the College of Engineering, Michigan State University, under the
Global Impact Initiative. G.A. and W.A.S. were supported by National Science Foundation under Award No. PHYS-1535279. The authors would like to thank Robert Rechenberg (Fraunhofer Center for Coatings and Diamond Technologies) for providing us with HPHT diamond sample for measurements. We would also like to thank Shengyuan Bai and Elias Garratt (Michigan State University) for assisting with GIXRD measurements.

\bibliography{antiDS}

\end{document}